\documentclass[aps,preprint]{revtex4}%
\usepackage{amsfonts}
\usepackage{amsmath}
\usepackage{amssymb}
\usepackage{graphicx}%
\providecommand{\U}[1]{\protect\rule{.1in}{.1in}}

\begin{document}
\preprint{ }
\title{Optimal Nonadditive Quantum Error-Detecting Code}
\author{Wen-Tai Yen}
\author{Li-Yi Hsu}
\affiliation{Department of Physics, Chung Yuan Christian University, Chungli 32081, Taiwan,
Republic of China}

\begin{abstract}
In this paper, we investigate the optimal nonadditive quantum error-detecting
codes with distance two. The the numerical simulation shows that, with $n$
being can be 5, 6, 7, 8, 10 and 12, such the $n$-qubit quantum error-detecting
codes with maximal number of codewords can be found. Therein, except the $n$=7
case, the $n$-vertex loop graphs help find the optimal quantum codes.

\end{abstract}
\volumeyear{year}
\volumenumber{number}
\issuenumber{number}
\eid{identifier}
\date[Date text]{date}
\received[Received text]{date}

\revised[Revised text]{date}

\accepted[Accepted text]{date}

\published[Published text]{date}

\maketitle

Quantum computation, with the use of the quantum mechanical phenomena, is
theoretically proved more efficient than classical computation on problems,
such as prime factorization and unsorted database search. However, in physical
realization, disturbance from environment can cause unavoidable errors in any
quantum information processing. To protect the quantum data, the quantum
error-correction/detection codes are exploited to defy decoherence. Over the
past decade, extensive studies on quantum error-correcting codes have enhanced
the feasibility of large-scale quantum computers in the foreseen future.
Therein, an important class of quantum error-correction codes are stabilizer
codes, which are additive.

Recently, nonadditive quantum codes without stabilizer structure have
attracted some attentions \cite{9710031,9703002}. In brief, an (($n$, $K$,
$d$)) nonadditive code indicates encoding a $K$-dimension subspace, which is
also called codespace, with distance $d$ into $n$ physical qubits. Smolin
\textit{et al }originally proposed the ((5, 6, 2)) nonadditive quantum
error-detection codes \cite{0701065}. Later a unifying approach was presented
by Cross \emph{et al }to construct additive and nonadditive codes, called
codeword stabilized codes \cite{07081021}. Yu \emph{et al} proposed the first
((9, 12, 3)) quantum code \cite{SixiaPRL}. Notably, so far almost nonadditive
quantum codes are on the graph-state basis. Some further properties of
codeword stabilized codes were investigated by Chuang \emph{et al
}\cite{0803.3232, 0808.3086}.\emph{ }Later many graphical quantum codes,
binary or non-binary, were proposed \cite{0709.1780, nonbinary,0712.1979}. It
has also been shown that nonadditive quantum codes can outperform the additive
ones in some aspects \cite{preexist}.

In this paper, we focus on the codeword stabilized codes with distance two.
Rains proposed the upper bound of $K$ with $d=2$ \cite{9704043}: if $n$ is
even ($n=2m$),
\begin{equation}
K\leq4^{m-1};
\end{equation}
otherwise ($n=2m+1$),%
\[
K\leq4^{m-1}(2-\frac{1}{m}).
\]

In this paper, we numerically verify that some $n$-qubit nonadditive codeword
stabilized codes with $\ n=5,6,7,8$ 10 and 12, respectively, can reach the
upper bounds. Before proceeding further, we review codeword stabilizer codes.
In brief, an (($n$, $K$)) codeword stabilizer code can be described on the
basis of a specific $n$-qubit graph state with the associated undirected graph
$\mathcal{G}_{c_{1}}=(V$, $E)$, $\left\vert V\right\vert =n$. Denote the
neighboring vertex set of the vertex $i$ as $N(i)=\{j|$ ($i$, $j$) $\in E\}$.
The $n\times n$ symmetric adjacency matrix with vanishing diagonal entries is
denoted by $\Gamma$ and the entry $\Gamma_{ij}=1$ if ($i$, $j$) $\in E$ and 0
otherwise. The graph state associated with the graph $\mathcal{G}_{c_{1}}$
reads
\[
\left\vert c_{1}\right\rangle =%
{\displaystyle\prod\limits_{(i,j)\in E}}
\mathcal{Z}_{ij}\left\vert +\right\rangle ^{\otimes n},
\]
where $\mathcal{Z}_{ij}$ is the controlled-phase operation between qubits
$i$\ and $j$. The corresponding density matrix can be expressed as
\begin{equation}
\rho_{c_{1}}=\frac{1}{2^{n}}%
{\displaystyle\prod\limits_{i=1}^{n}}
(\mathbf{I}+g_{i}),
\end{equation}
where $\mathbf{I}$ is the identity matrix and the stabilizer generator $g_{i}$
can be written as
\begin{equation}
g_{i}=X_{i}%
{\displaystyle\prod\limits_{j\in N(i)}}
Z_{j},\text{ }i=1,\cdots n, \label{g}%
\end{equation}
where $X_{i}$\ and $Z_{j}$\ are Pauli matrices $\sigma_{x}$ and $\sigma_{z}$
on the qubit $i$ and $j$, respectively. It is noteworthy that
\begin{equation}
\{Z_{i}\text{, }g_{i}\}=0 \label{z}%
\end{equation}
The codespace of the optimal quantum error-detection code, denoted by
$\mathfrak{C}_{Q}$ , with distance two are spanned by the orthonormal state
set $\mathcal{C}_{set}=$\{$\left\vert c_{L}\right\rangle |L=1,\cdots,K_{\max}%
$\}, where%
\begin{equation}
\left\vert c_{L}\right\rangle =%
{\displaystyle\prod\limits_{j=1}^{n}}
Z_{j}^{b_{L_{j}}}\left\vert c_{1}\right\rangle ,b_{L_{j}}\in\{0,1\}, \label{l}%
\end{equation}
where $K_{\max}$ is equal to $2^{n-2}$\ or $\lfloor2^{n-2}(1-\frac{1}%
{n-1})\rfloor$ if $n$ is even or odd, respectively. The density matrix of
$\left\vert c_{L}\right\rangle $, $\rho_{c_{L}}$, is
\begin{equation}%
{\displaystyle\prod\limits_{j=1}^{n}}
Z_{j}^{b_{L_{j}}}\rho_{c_{L}}Z_{j}^{b_{L_{j}}}=\frac{1}{2^{n}}%
{\displaystyle\prod\limits_{j=1}^{n}}
(\mathbf{I}+(-1)^{b_{L_{j}}}g_{j}). \label{b}%
\end{equation}
In this paper, we character $\left\vert c_{L}\right\rangle $ by the
eigenvalues of operators $g_{1}$, $g_{2}$, $\cdots$ , $g_{n}$.\ That is, the
$n$-bit string $c_{L}$ are expressed as
\begin{equation}
b_{L_{1}}b_{L_{2}}\cdots b_{L_{n}}, \label{re}%
\end{equation}
since, as according to Eq. (\ref{b}), the eigenvalue of the $g_{j}$ the
eigenstate $\left\vert c_{L}\right\rangle $ is \bigskip\ $b_{L_{j}}$
($c_{1}=0^{\otimes n}$).

The essential advantage of codeword stabilized codes lies on its
correspondence to a classical error-correction/detection code with a specific
error model. To see this, in the corresponding classical code of
$\mathfrak{C}_{Q}$, denoted by $\mathfrak{C}_{C}$, the $n$-bit codeword set is
$\mathcal{C}_{classical}=$\{$c_{L}|L=1,\cdots,K_{\max}$\}. The codeword state
$\left\vert c_{L}\right\rangle $ corresponds the $n$-bit codeword $c_{L}$. Now
let $\left\vert c_{L}^{\overline{i}}\right\rangle =Z_{i}\left\vert
c_{L}\right\rangle .$ According to Eqns. (\ref{b}) and (\ref{re}),
$c_{L}^{\overline{i}}=b_{L_{1}}b_{L_{2}}\cdots b_{L_{i-1}}\overline{b_{L_{i}}%
}b_{L_{i+1}}\cdots b_{L_{n}}$. That is, in the quantum codespace, the
phase-flip error on qubit $i$ corresponds to the bit-flip error on bit $i$ in
the classical codespace. Furthermore, according to Eq. (\ref{g}), it is
obvious to verify that%

\begin{equation}
X_{i}\left\vert c_{L}\right\rangle =%
{\displaystyle\prod\limits_{j\in N(i)}}
Z_{j}\left\vert c_{L}\right\rangle \label{flip}%
\end{equation}
The bit-flip error on the $i$-th qubit corresponds the multi-qubit phase-flip
errors on the neighboring qubits in the quantum codespace. As a result, all
single-qubit errors can be regards to only phase-flip errors, which
corresponds bit-flip errors on one or more bits in the corresponding classical
error model.

In details, the phase-flip error $Z_{i}$ corresponding bit-flip on the $i$-th
bit. (In the following, by $\overline{1_{k}}$, we denote $n$-bit string, where
$k$-th bit is 1 and the other $n-1$ bits are zeros.) In addition, according to
Eq. (\ref{flip}), the single-qubit bit-flip error $X_{i}$ corresponds to
classical $\left\vert N(i)\right\vert $ bit-flip errors on bits $j_{1}%
$,$\cdots$, $j_{N(i)}$, where $(i,$ $j_{k})\in E$. As an example, Cross
\emph{et al} have originally proposed the optimal (($5$, $6$, 2)) codeword
stabilized code, where the 5-qubit graph states with 5-vertex cycle (or loop)
graph are exploited. Therein, five single-qubit bit-flip errors corresponds to
the two-bit-flip classical errors, which are 10100, 01010, 00101, 10010 and
01001, respectively. Finally, the corresponding classical error of
single-qubit error $Y_{i}$ corresponds flipping $(\left\vert N(i)\right\vert
+1)$ bits $j_{1}$,$\cdots$, $j_{|N(i)|}$, and $i$. In the following, the
$n$-bit strings $\Gamma_{i1}\Gamma_{i2}\cdots\Gamma_{i(i-1)}0\Gamma
_{i(i+1)}\cdots\Gamma_{in}$ and $\Gamma_{i1}\Gamma_{i2}\cdots\Gamma
_{i(i-1)}1\Gamma_{i(i+1)}\cdots\Gamma_{in}$ by $\Gamma_{i}^{0}$ and
$\Gamma_{i}^{1}$, respectively.

Here we define the state set $\mathcal{C}_{i,k}^{Q}=\{\left\vert
c_{i}\right\rangle $, $Z_{k}\left\vert c_{i}\right\rangle ,$ $X_{k}\left\vert
c_{i}\right\rangle $, $Y_{k}\left\vert c_{i}\right\rangle \}$, where $1\leq
k\leq n$. Obviously,
\begin{equation}
\mathcal{C}_{i,k}^{Q}\cap\mathcal{C}_{j,k}^{Q}=\varnothing\text{ }\forall
i\neq j. \label{set}%
\end{equation}
The corresponding $n$-bit string set is denoted by $\mathcal{C}_{i,k}=$
\{$c_{i}$, $c_{iz}=c_{i}+\overline{1_{k}}$, $c_{ix}=c_{i}+\Gamma_{k}^{0}$,
$c_{iy}=c_{i}+\Gamma_{k}^{1}$\}. Also, $\forall i\neq j$, $\mathcal{C}%
_{i,k}\cap\mathcal{C}_{j,k}=\varnothing$.

However, the associated graph of the graph state $\left\vert c_{1}%
\right\rangle $ and hence the corresponding error strings $\Gamma_{i}^{0}$ and
$\Gamma_{i}^{1}$ are unknown. For a given graph $\mathcal{G}_{c_{1}}^{\prime}%
$as the associated graph of the graph state $\left\vert c_{1^{\prime}%
}\right\rangle $, we test whether there is a error-detection code,
$\mathcal{C}_{classical}$, with distance two. Here we brief our algorithm as follows.

(i) Given a $\mathcal{G}_{c_{1}}^{\prime}$and the corresponding error strings
$\Gamma_{i}^{0\prime}$ and $\Gamma_{i}^{1\prime}$ as inputs. Set the value of
the variable count as 0.

(ii) Generate $2^{n-2}$ $n$-bit binary strings $s_{1},\cdots,s_{2^{n-2}},$
($s_{1}=0^{\otimes n}$) as if $\left\vert s_{1}\right\rangle ,\cdots
,\left\vert s_{2^{n-2}}\right\rangle $ were codeword states.

(iii) Define the $2^{n-2}$ 4-element string set $S_{1},\cdots,S_{2^{n-2}}$,
where $S_{i}=\{s_{i},s_{ix},s_{iz},s_{iy}\}$ and $s_{iz}=s_{i}+\overline
{1_{1}}$, $s_{ix}=s_{i}+\Gamma_{1}^{0\prime}$ and $s_{iy}=s_{i}+\Gamma
_{1}^{1\prime}$. Then verify whether the condition
\begin{equation}
S_{i}\cap S_{j}=\varnothing\text{, }\forall i\neq j \label{criterion}%
\end{equation}
is satisfied. If not, go to (ii), else count=count+1 and do (iv).

(iv) Let $c_{1}^{\prime}=0^{\otimes n}$ and $c_{i}^{\prime}\in S_{i}$ $1\leq
i\leq2^{n-2}$. Check whether the set $\mathcal{C}^{\prime},$
\begin{equation}
\mathcal{C}^{\prime}=\{c_{i}^{\prime}|c_{l}^{\prime}\neq c_{m}^{\prime
}+\overline{1_{k}}\wedge c_{l}^{\prime}\neq c_{m}^{\prime}+\Gamma_{k}%
^{0\prime}\wedge c_{l}^{\prime}\neq c_{m}^{\prime}+\Gamma_{k}^{1\prime}\text{,
}\forall k\text{, }l\text{, }m\text{ and }l\neq m\}, \label{cc}%
\end{equation}
where $1\leq l$, $m\leq2^{n-2}$ and $1\leq k\leq n$, exists. If yes,
$\mathcal{C}^{\prime}=\mathcal{C}_{classical}$ as the output, else if count
$\geq$ M we halt the program, else repeat (ii)-(iv).

In details, in the proposed algorithm, the set $S_{i}$ generated in (ii) is
presumed to be some set $\mathcal{C}_{j,1}$ and hence $c_{j}\in$ $S_{i}$ is
presumed. To validate the assumption, the Eq. (\ref{criterion}) as the
necessary condition is tested in (iii). If the assumption is not defied in
(iii), in (iv) $c_{i}^{\prime}$ is assumed to equal to $c_{j}$. If the
assumption is validated, there must be some set $\mathcal{C}^{\prime
}=\mathcal{C}_{classical}$. In addition, to generate the string set
$S=\{s_{i}|1\leq i\leq2^{n-2}\}$ in (ii), we random generate $n-2$
linear-independent $n$-bit string $x_{1},\cdots,x_{2^{n-2}}$. Then set $S$ we
exploited is $\{\sum_{i=1}^{n-2}\oplus b_{i}x_{i}|b_{i}\in\{0,1\}\}$.
Interestingly, such way of generating $S$ indeed helps find the optimal
error-detection codes. Moreover, for a given graph, we halt the program in
(iv)\ after $M$ failures of searching codewords.

Notably, the proposed algorithm is suitable for the even $n$ case, since the
upper bound is exactly equal to $2^{n-2}$. In this case, $\left\vert
\mathcal{C}_{classical}\right\vert =\left\vert S\right\vert $ and, as a
result, we just verify whether the set $C^{\prime}$ validates the condition in
Eq. (\ref{cc}). On the other hand, $\left\vert \mathcal{S}\right\vert
>\left\vert \mathcal{C}_{classical}\right\vert $ if $n$ is odd. Therefore, in
the odd $n$ case, the algorithm is modified as follows: In (iv), we replace
\{$c_{1}^{\prime}$, $c_{2}^{\prime}$, $\cdots$, $c_{2^{n-2}}^{\prime}$\} by
all of its $\lfloor2^{n-2}(1-\frac{1}{n-1})\rfloor$-element subset to check
whether Eq. (\ref{cc}) is satisfied. 

In our simulation, we explore the optimal $n$-qubit quantum error-detection
codes with distance two, where $5\leq n\leq12$. Figures (1-6) show the
associated graphs of some codeword states with $n$ being 5, 6, 7, 8, 10 and
12, respectively \cite{graph}.  In addition, it is noteworthy that, except the
$n=7$ case, the loop graphs are exploited to find the optimal quantum
error-detection codes with distance two. Finally, lthough a lot of 9- or
11-vertex graphs have been tried, we fail to find the optimal codes. 

The author LYH thanks to Dr. I-Ming Tsai for helpful discussion. He also
acknowledges support from National Science Council of the Republic of China
under Contract No. NSC.96-2112-M-033-007-MY3.

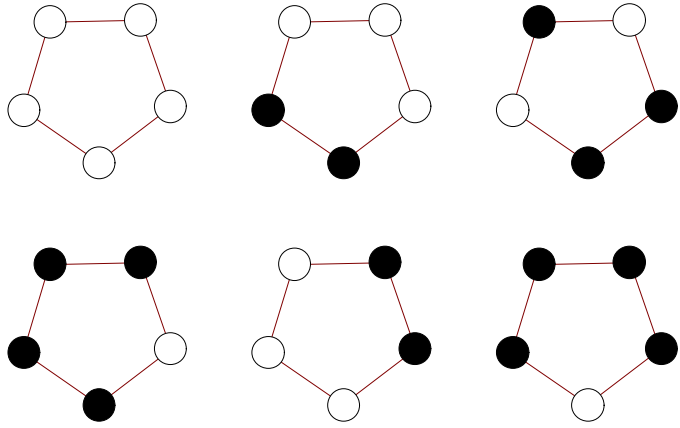
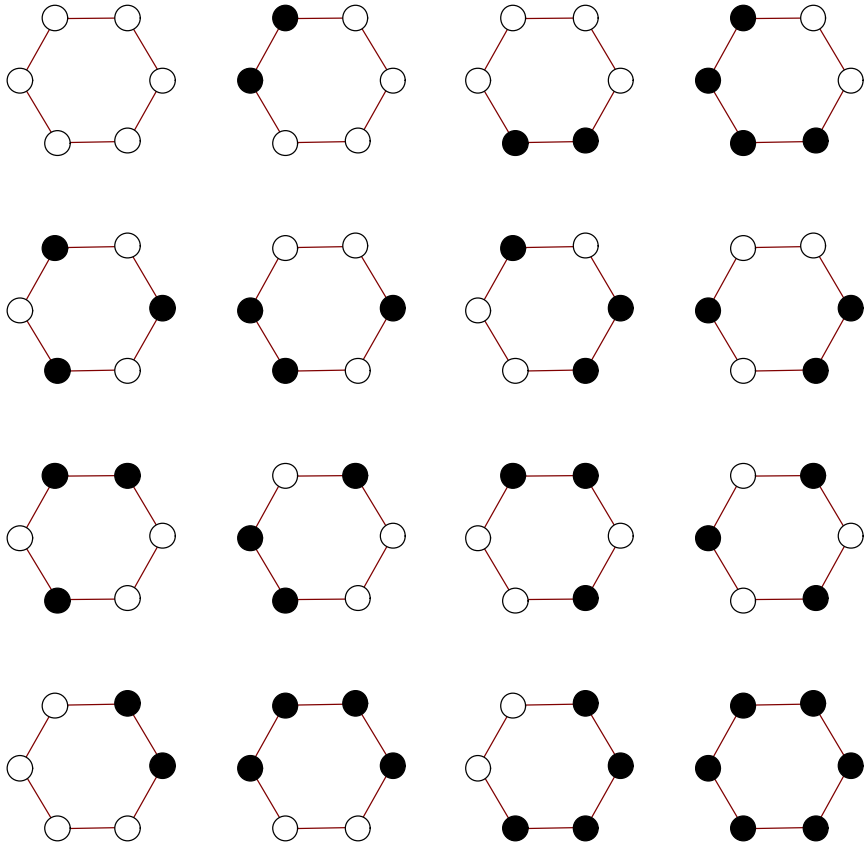
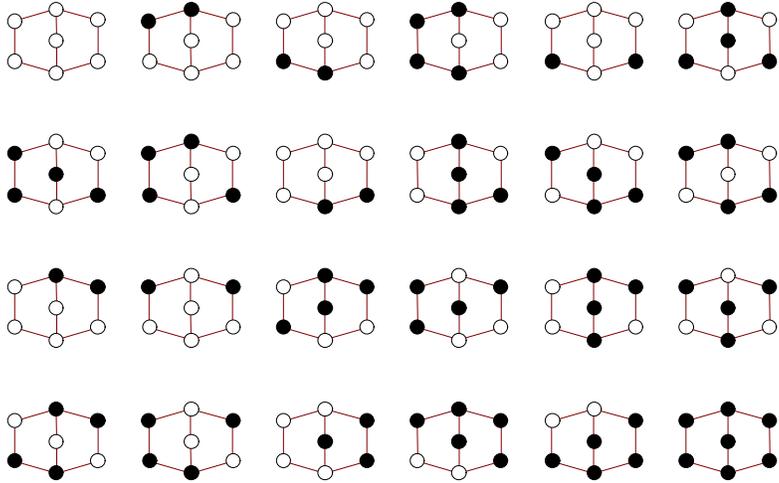
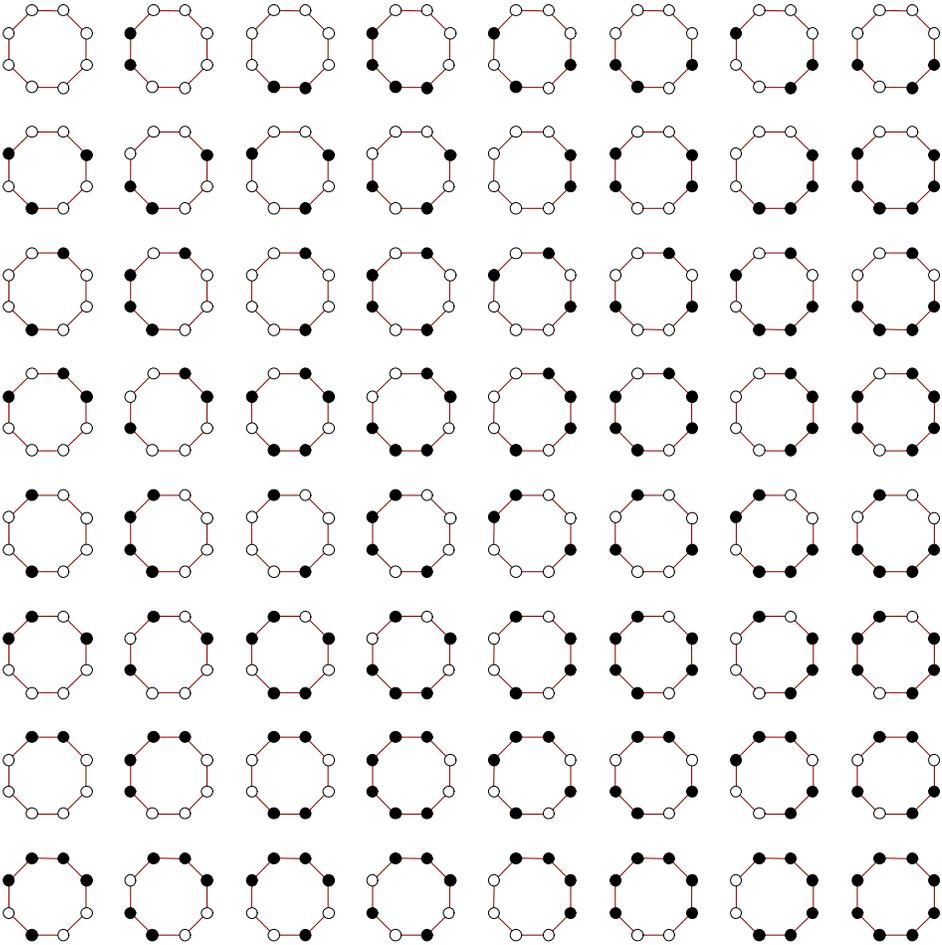
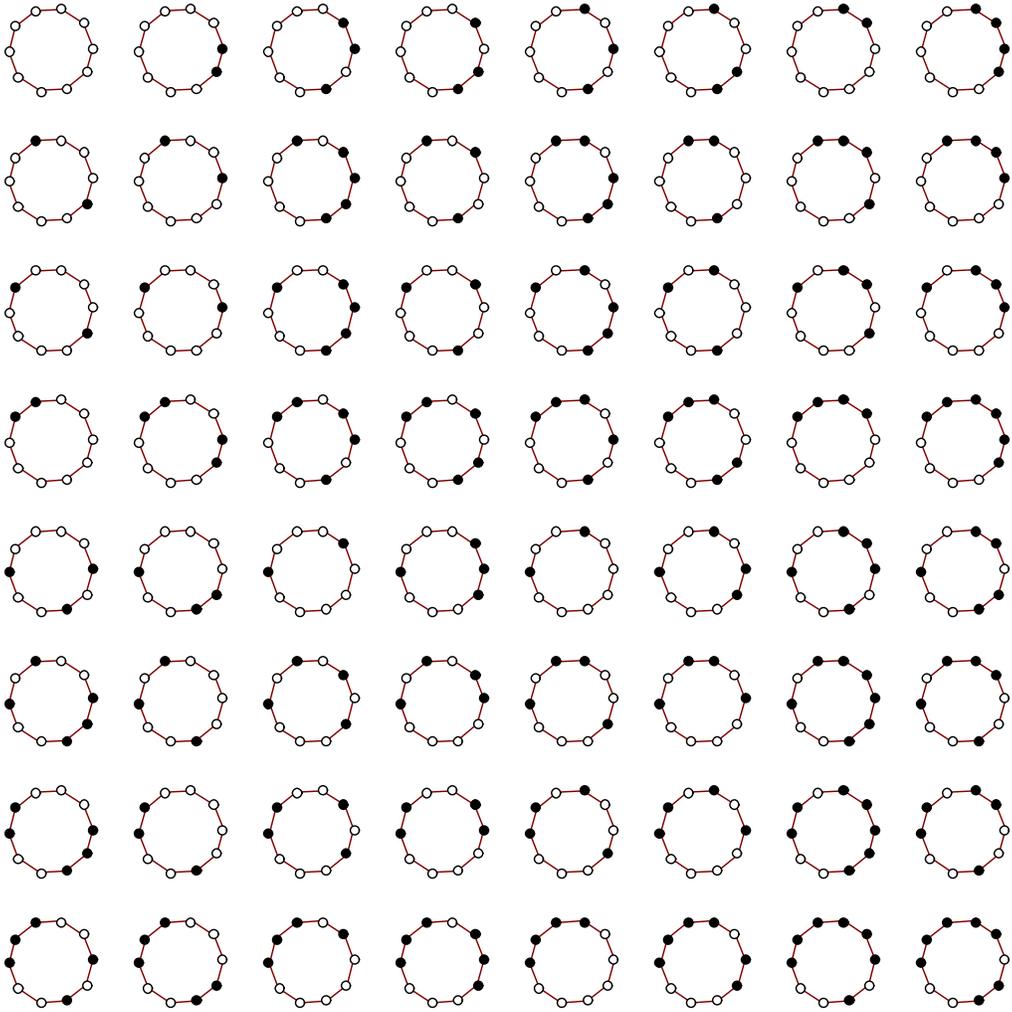
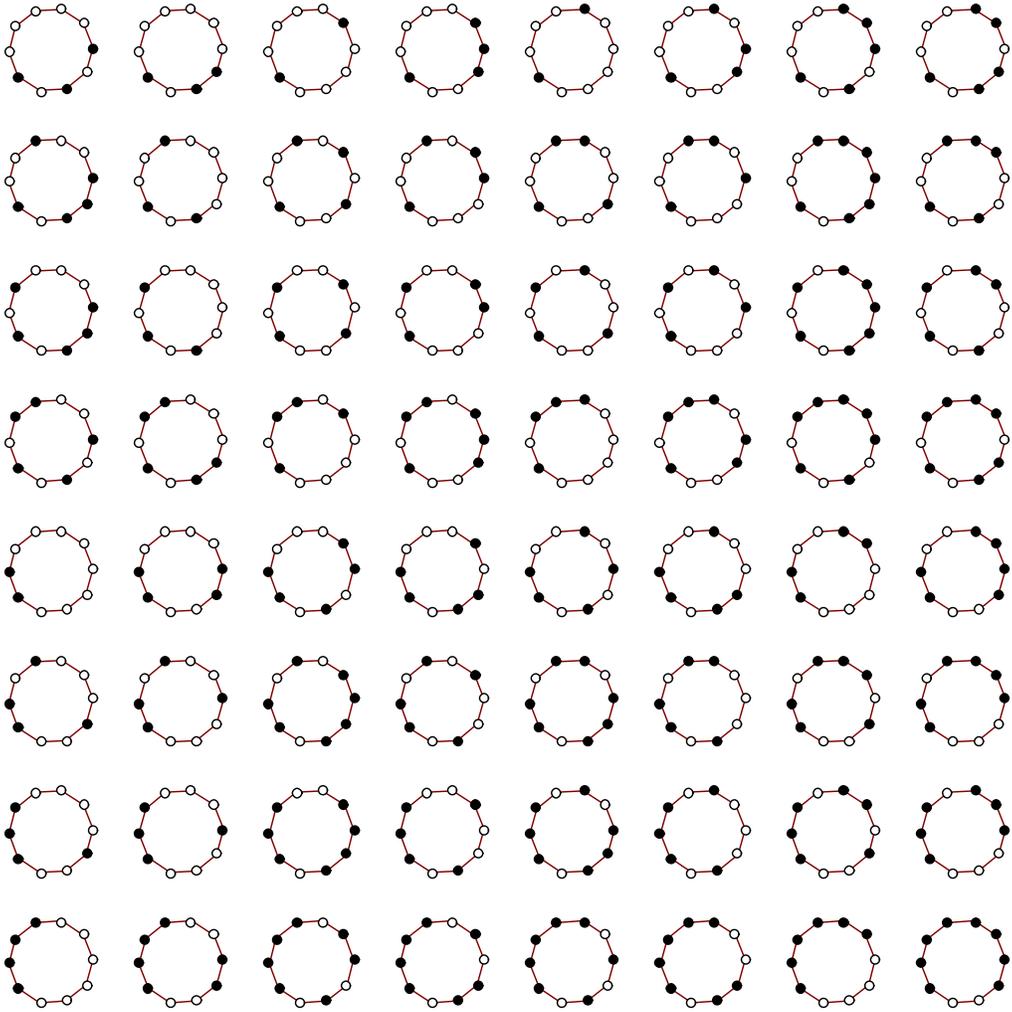
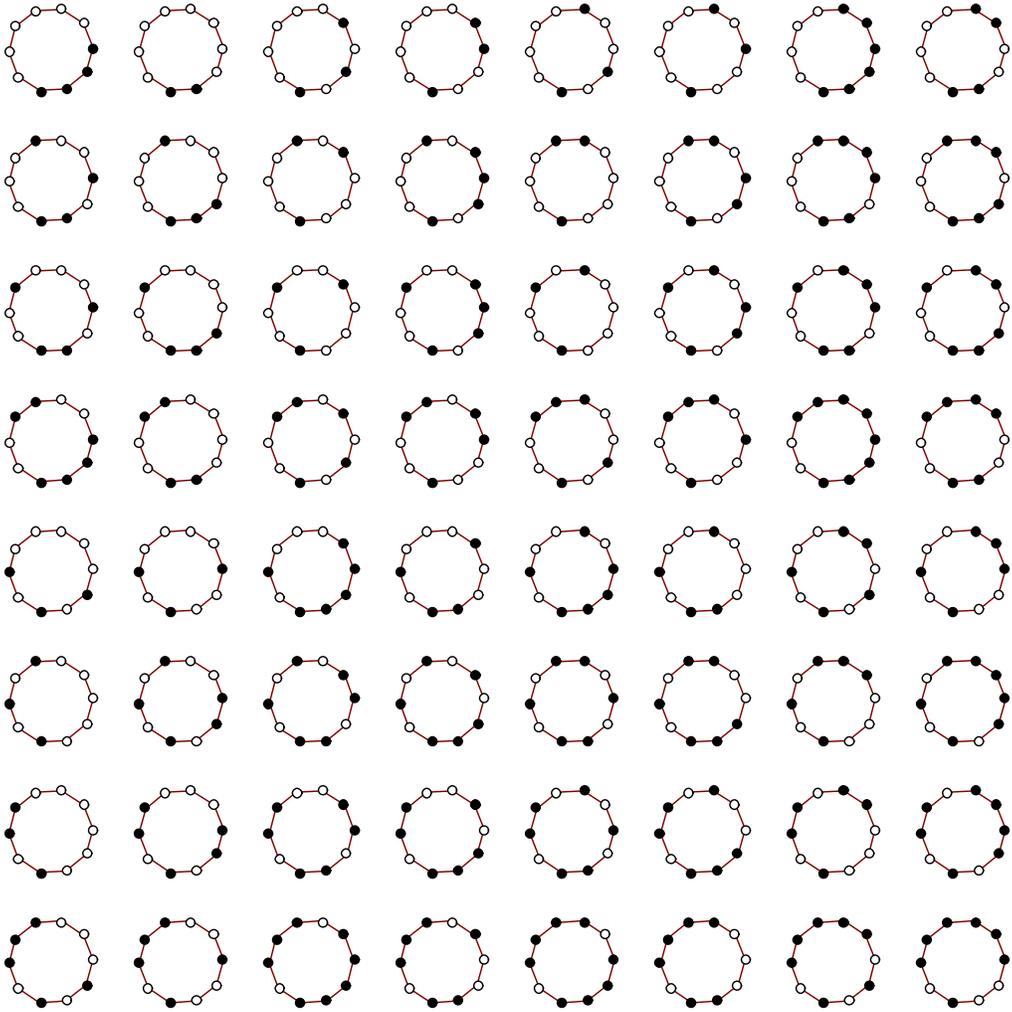
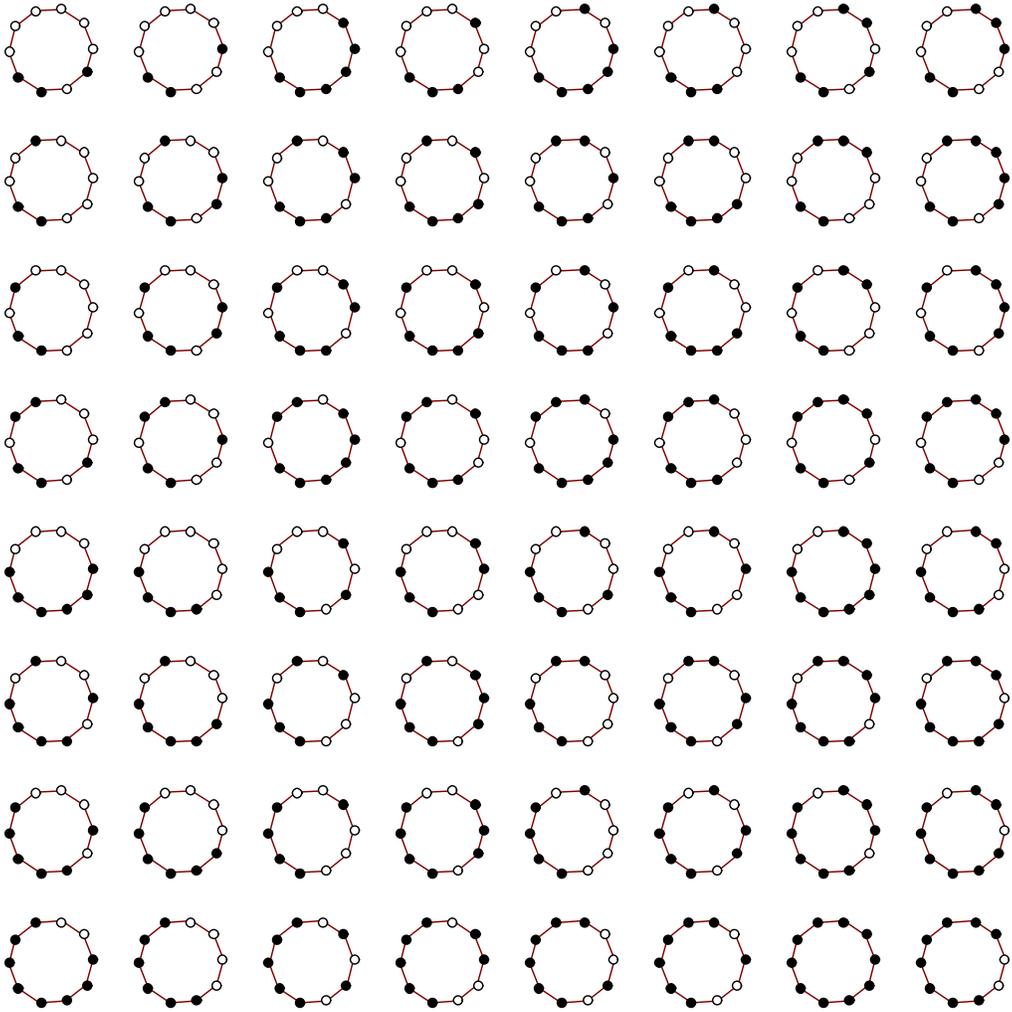
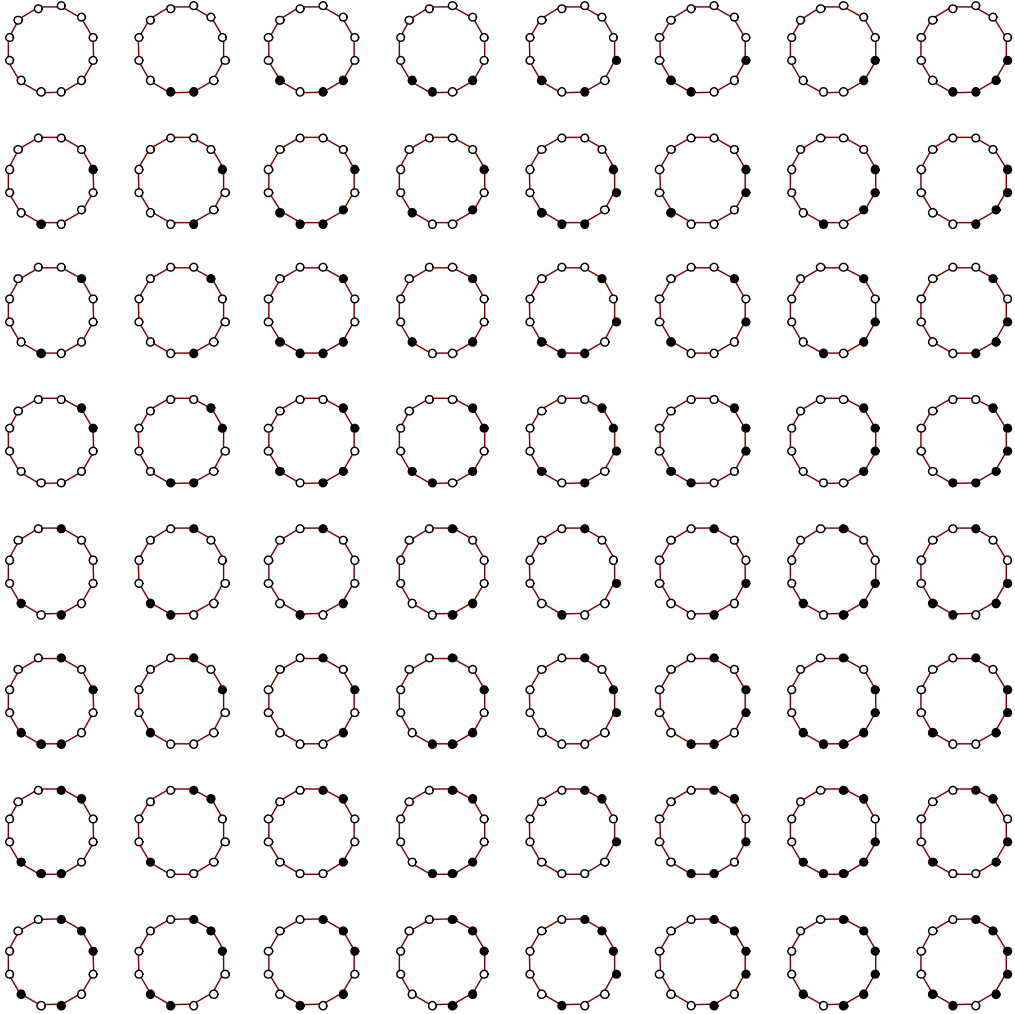
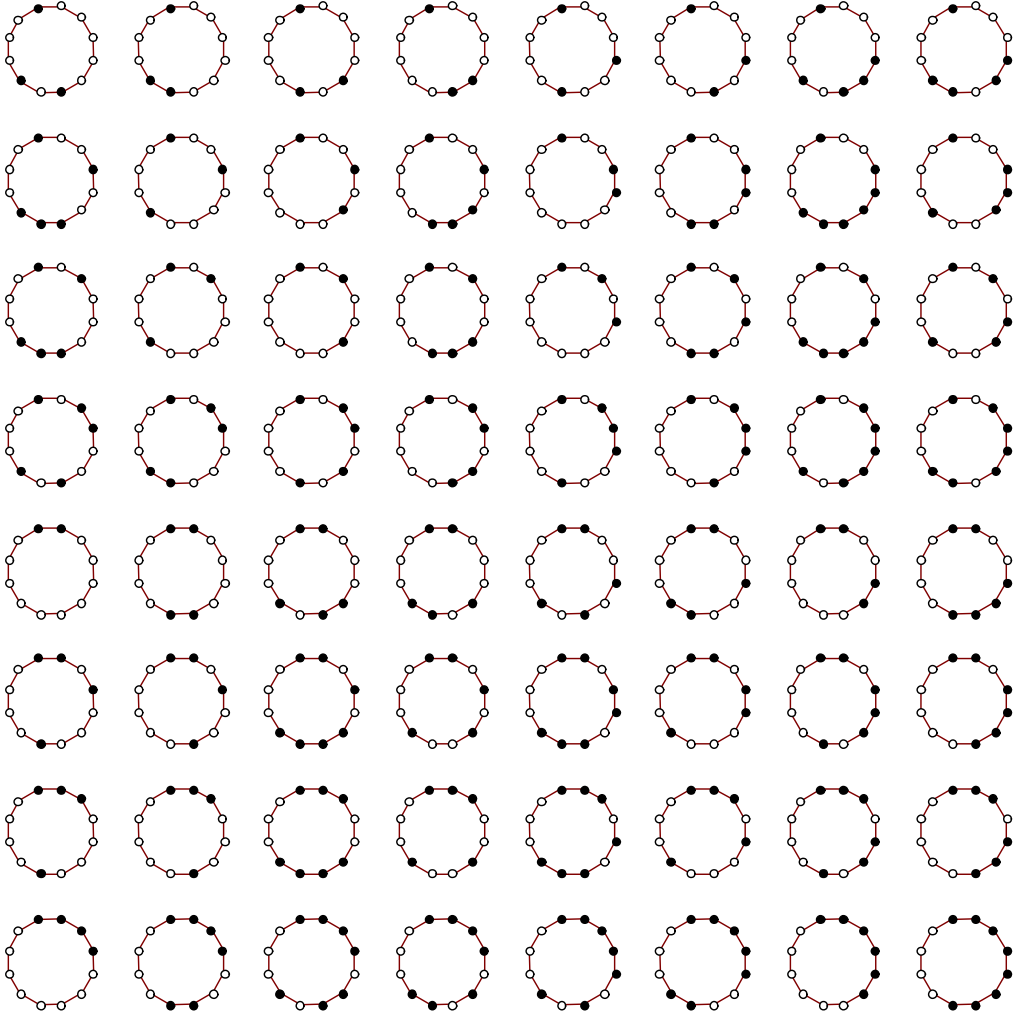
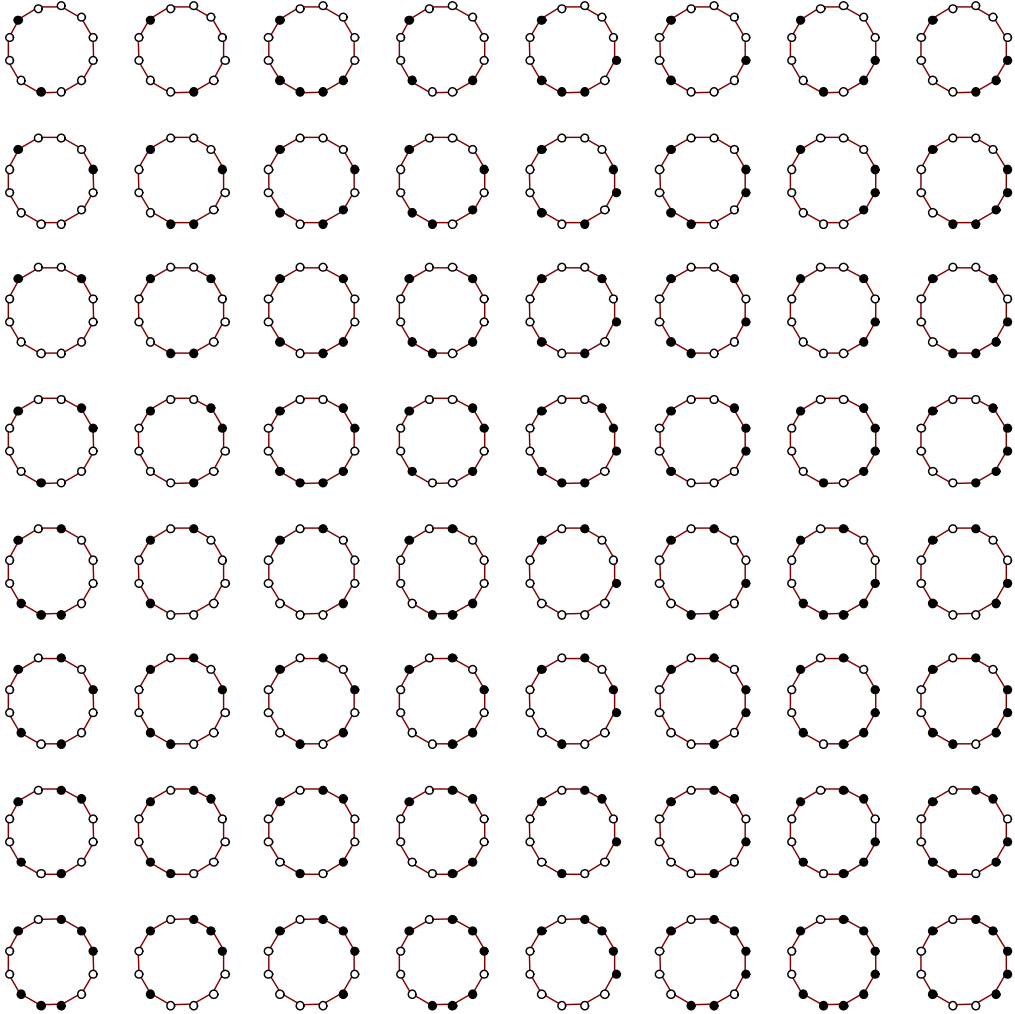
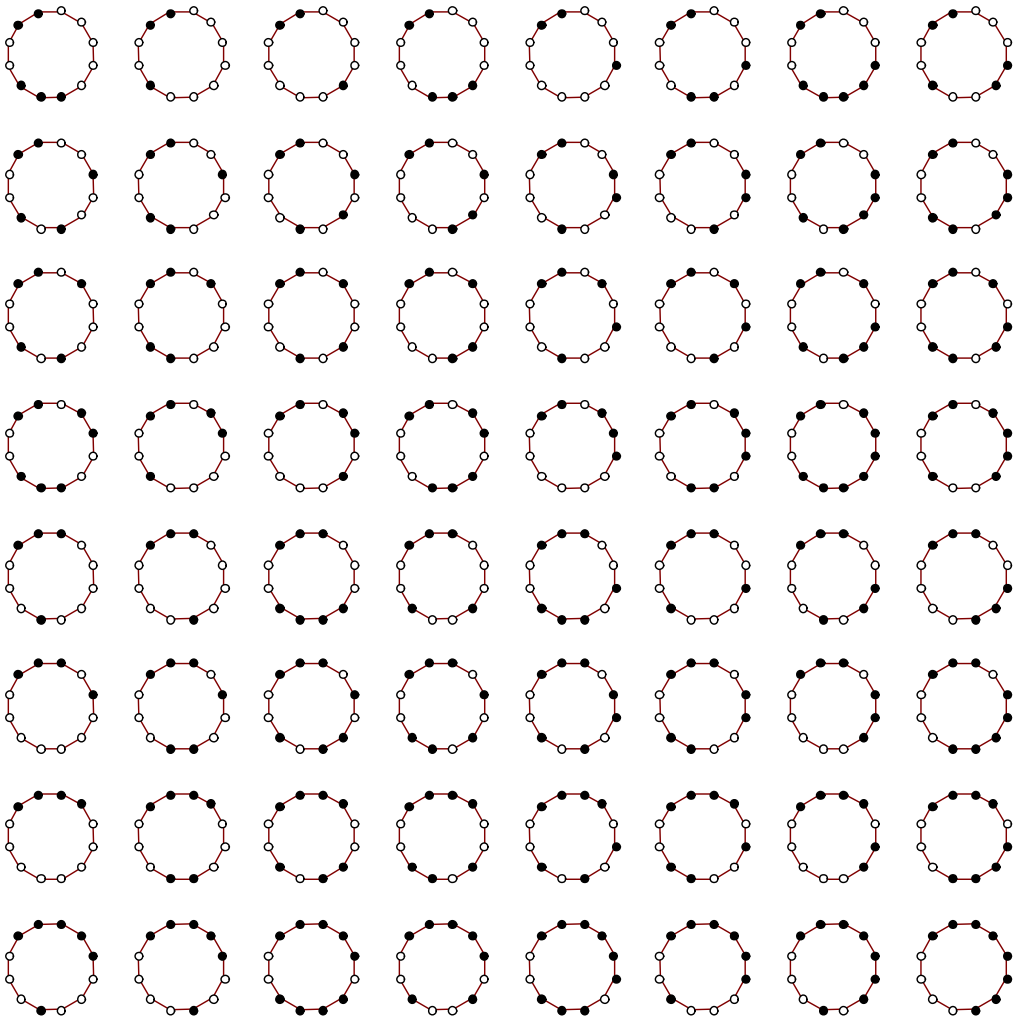
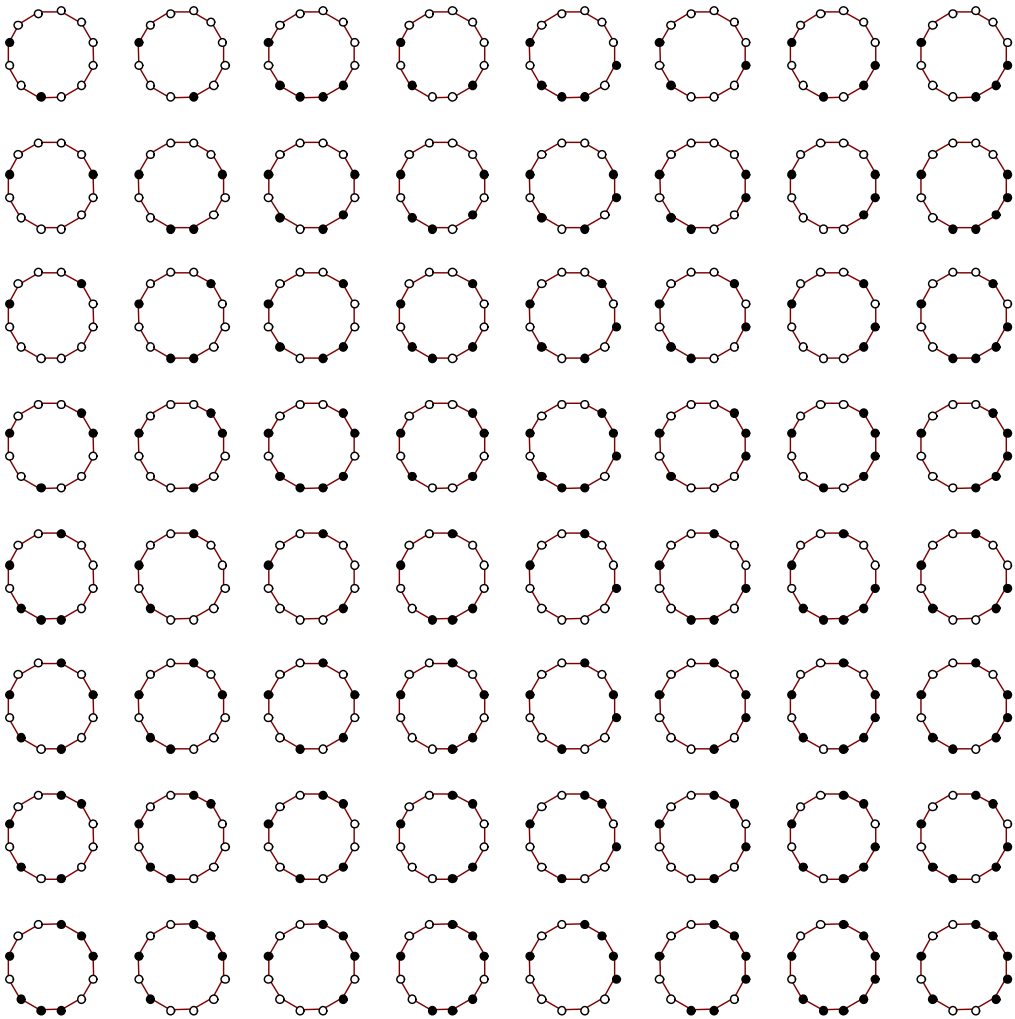
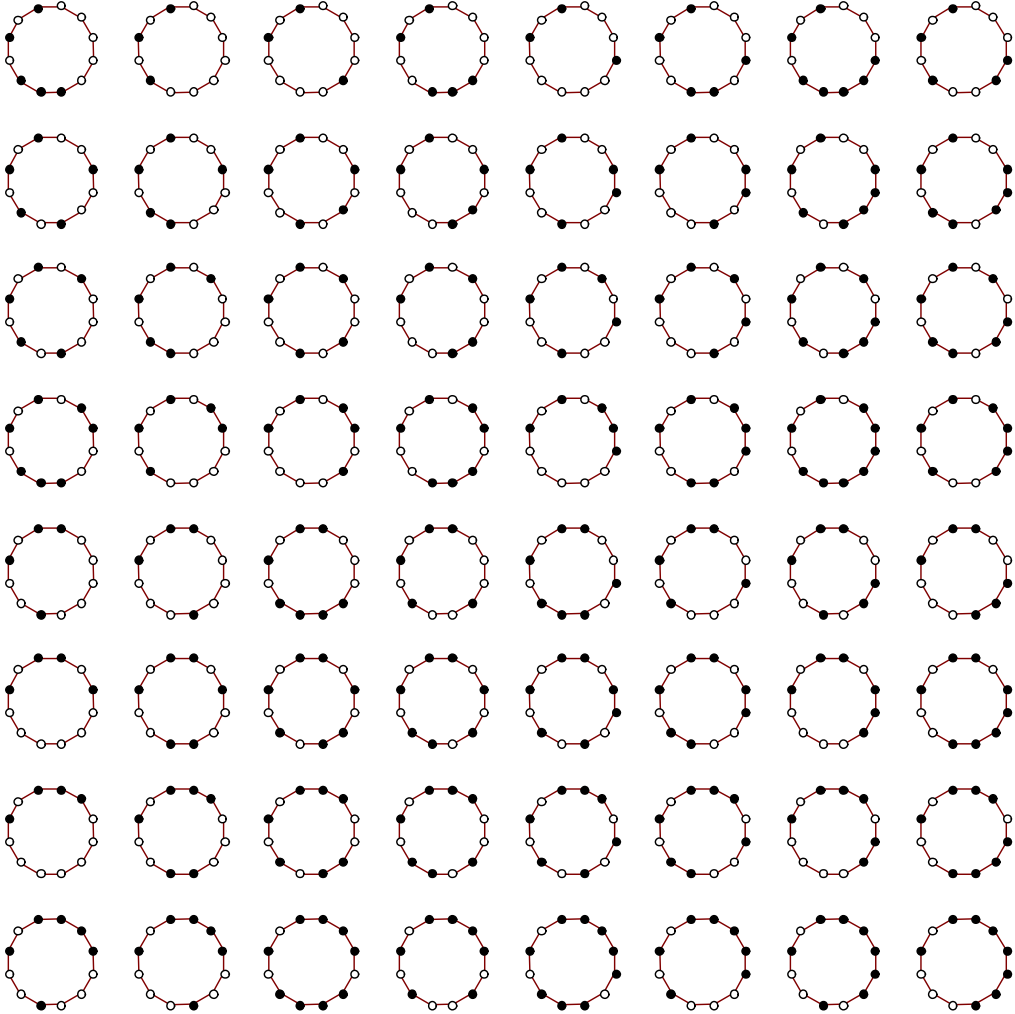
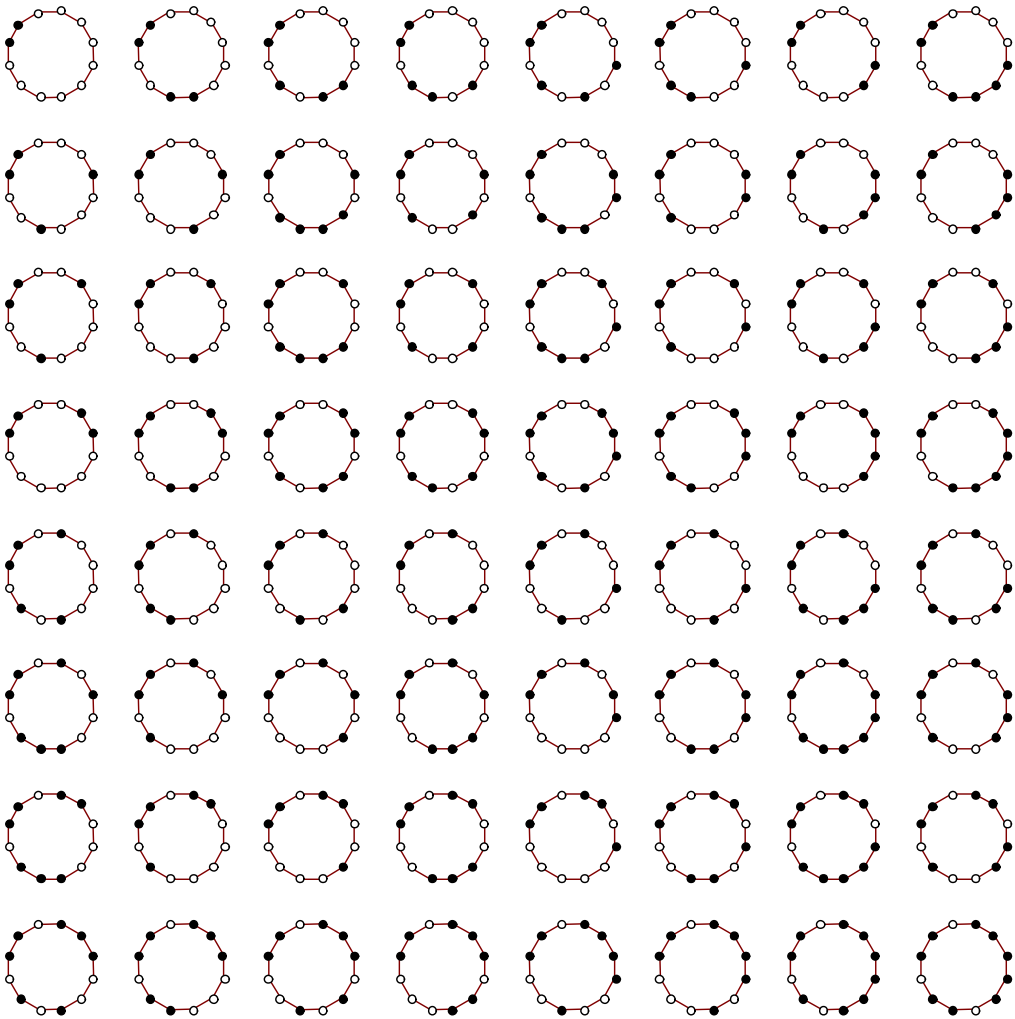
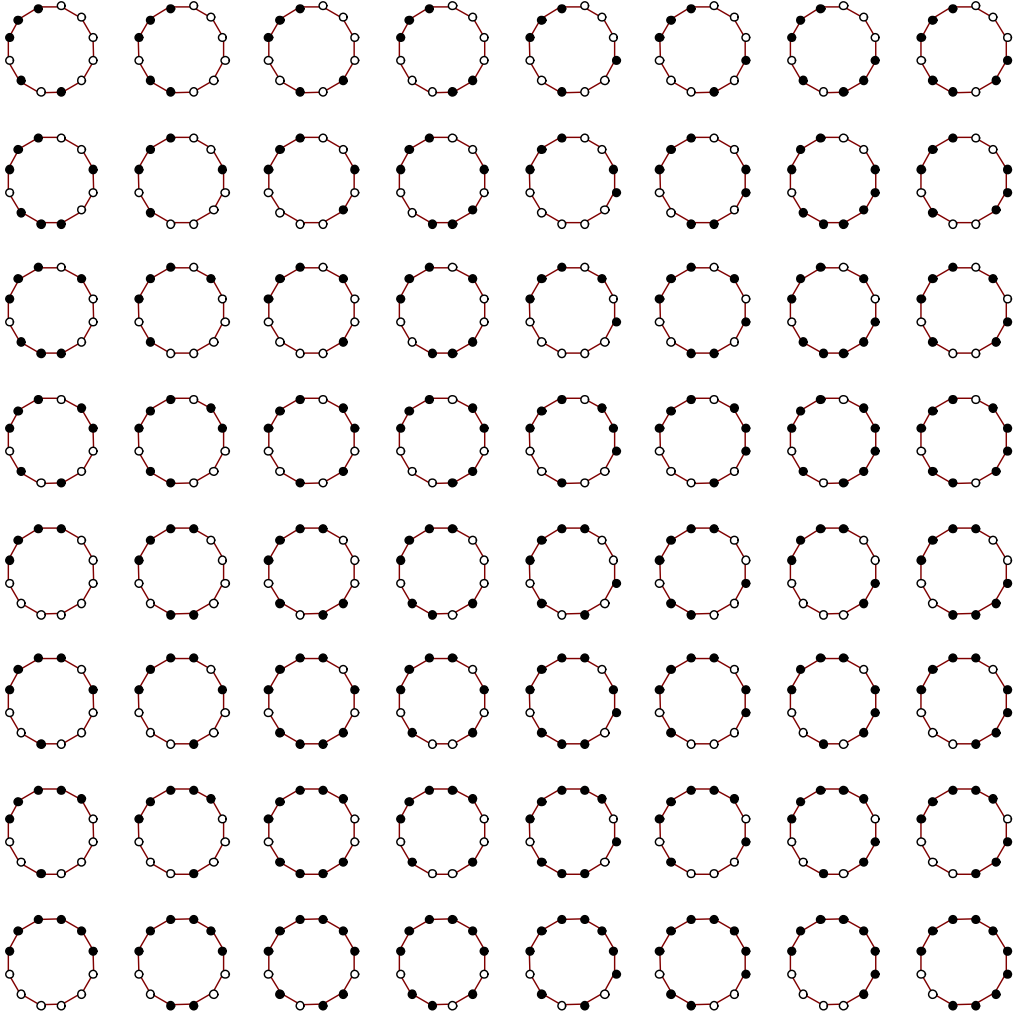
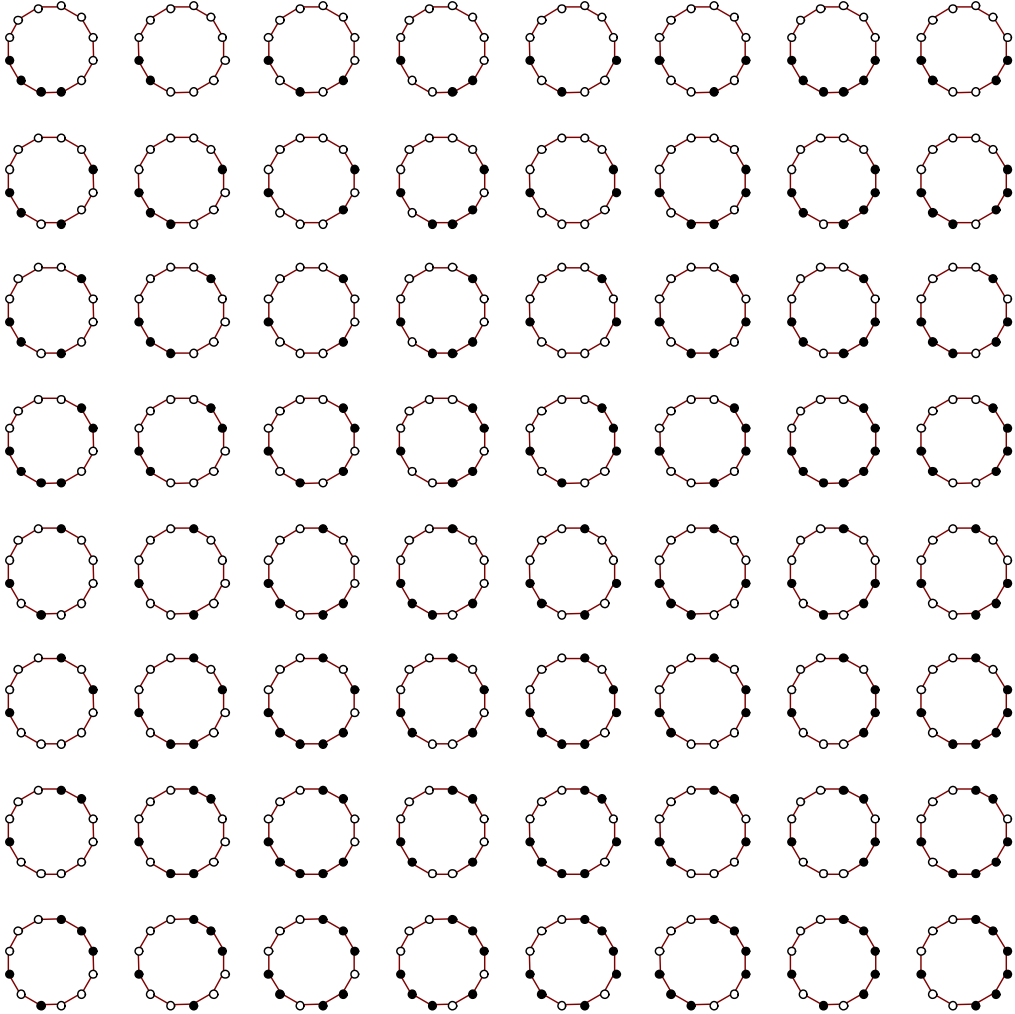
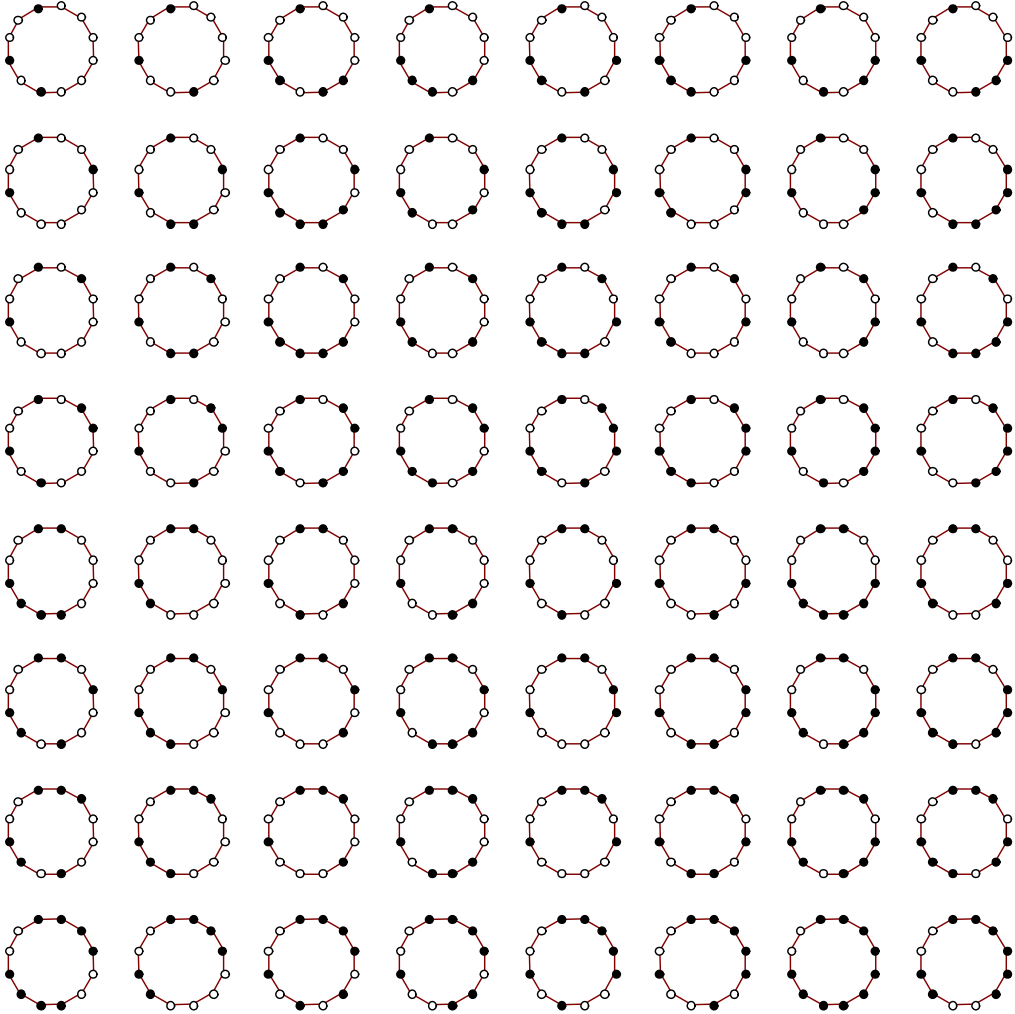
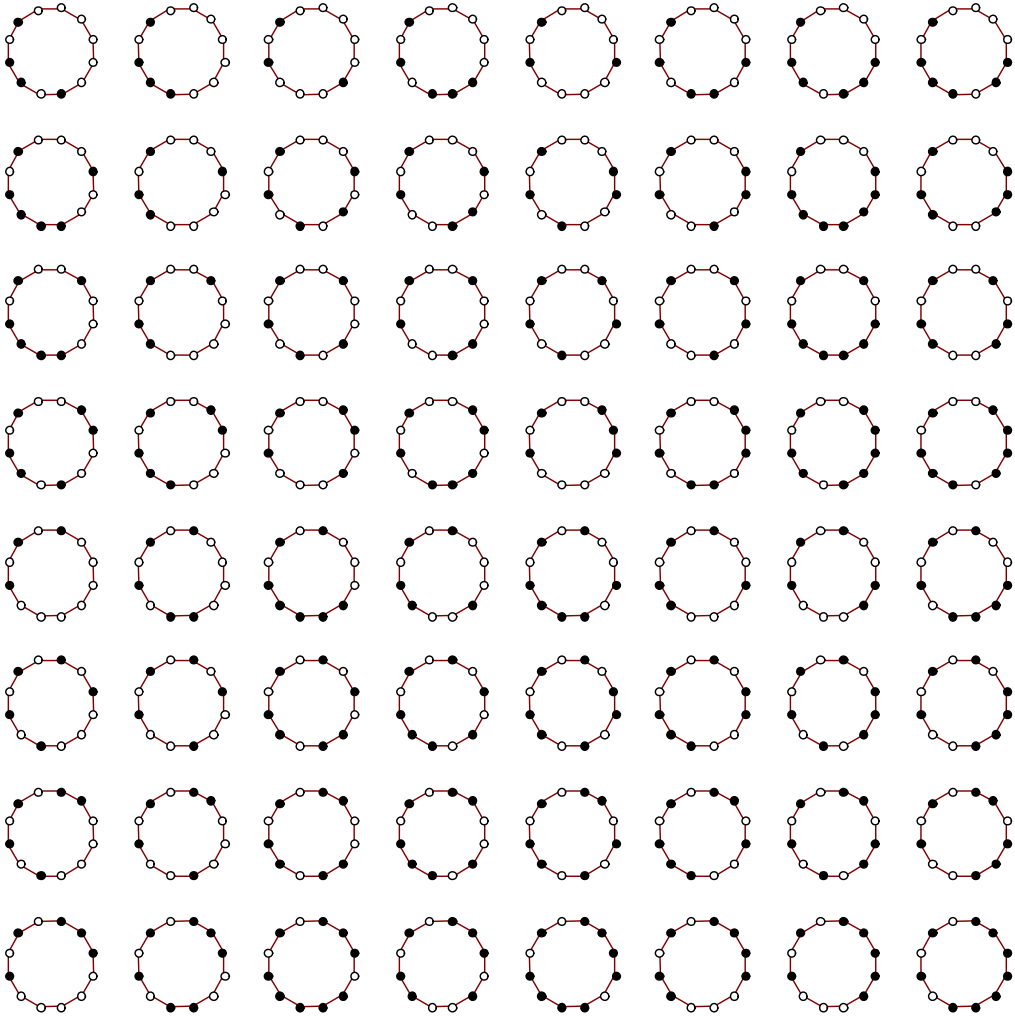
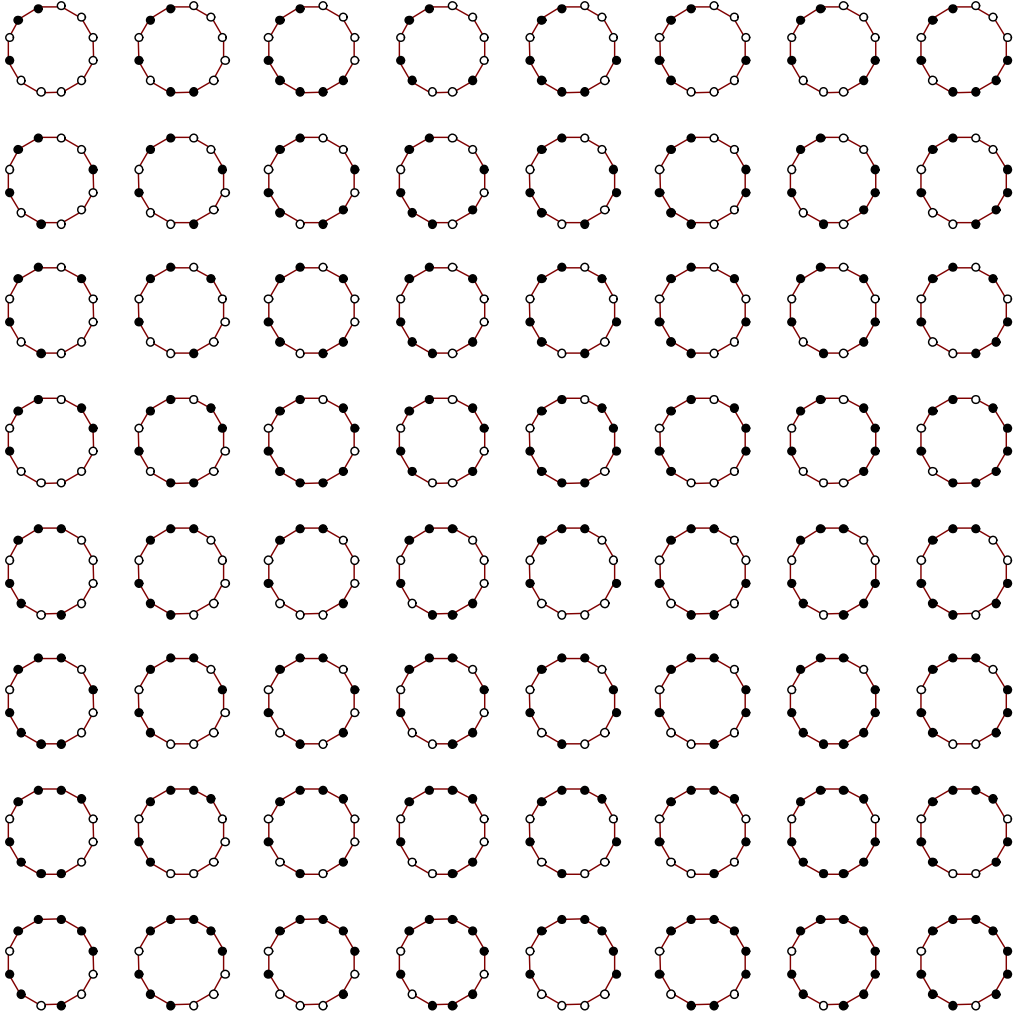
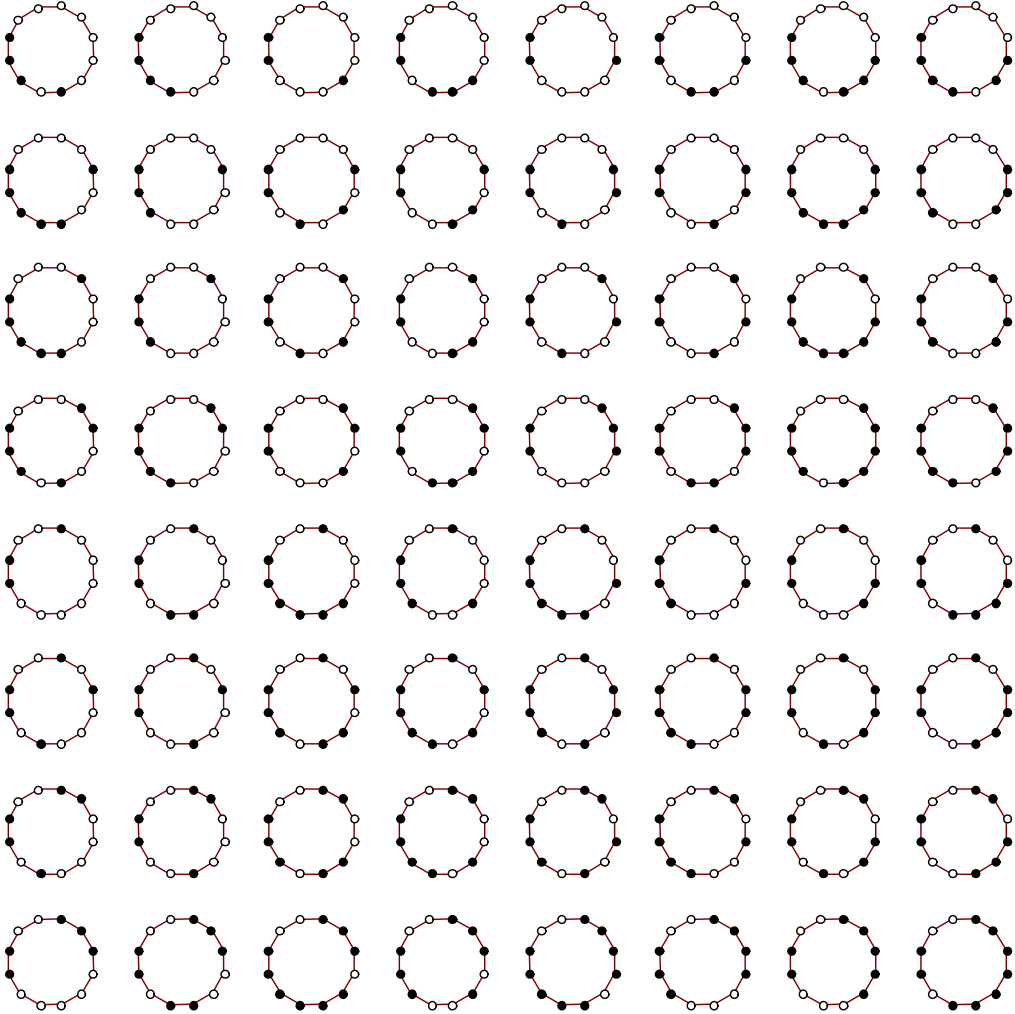
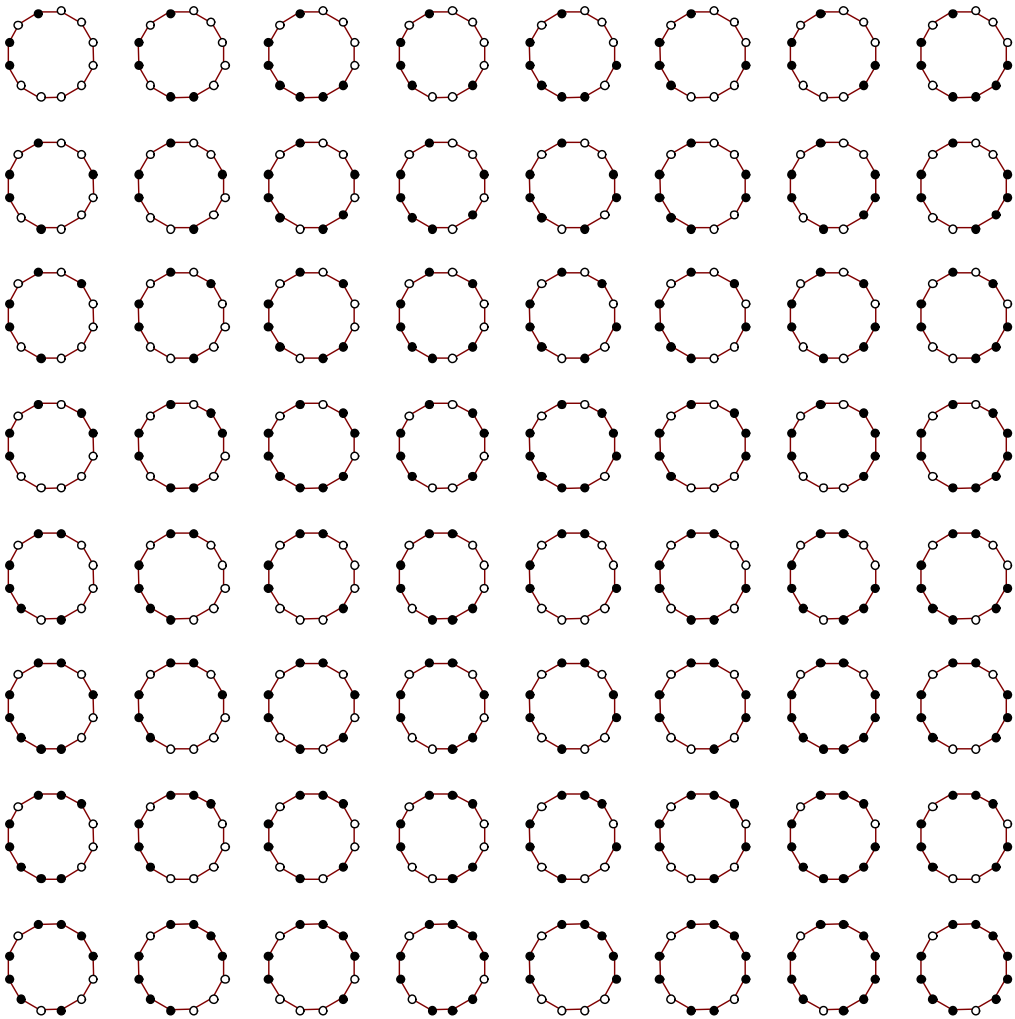
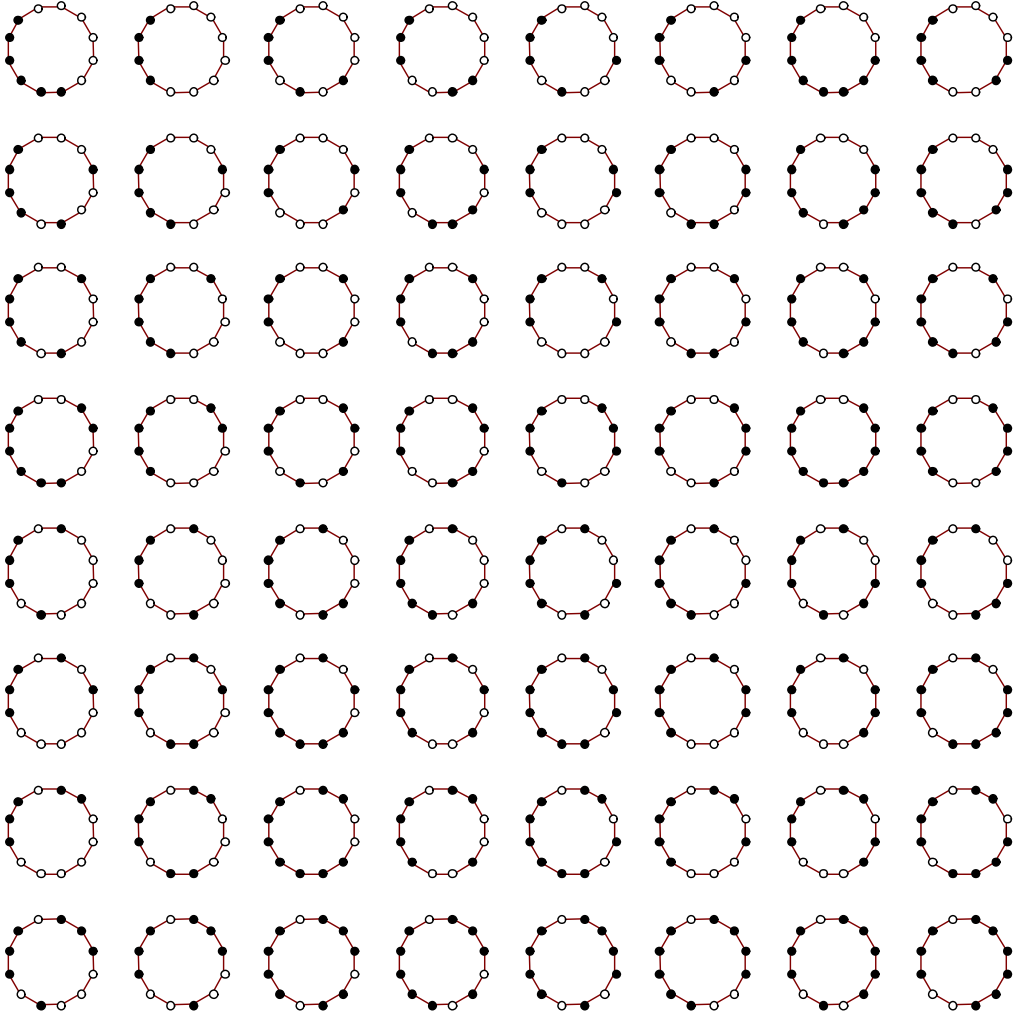
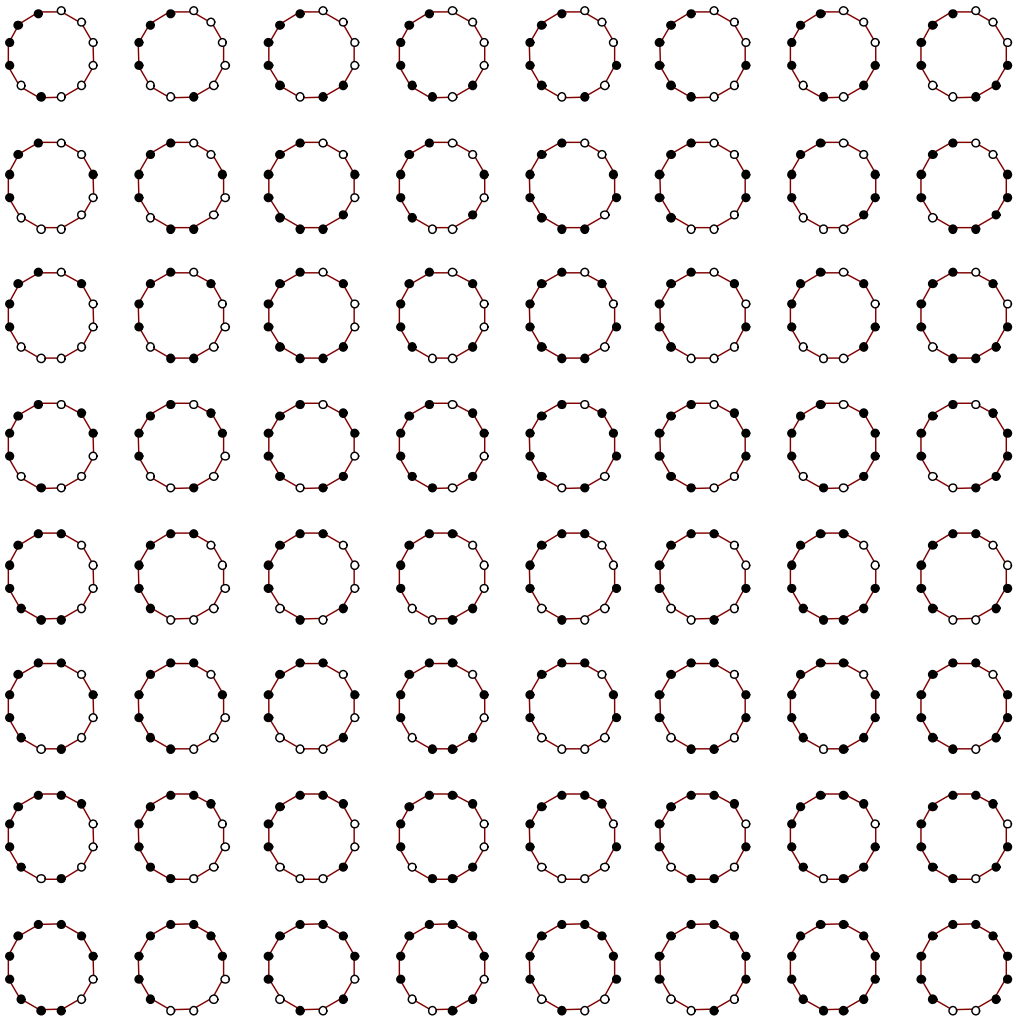
\end{document}